\documentclass[reprint,prl]{revtex4-1}
\usepackage{amsmath}
\usepackage{amssymb}
\usepackage{amsfonts}
\usepackage{natbib}
\newcommand{\p}{\partial}
\newcommand{\g}[1]{\mbox{\boldmath $#1$}}

\newcommand{\lp}{\left(}
\newcommand{\rp}{\right)}

\newcommand{\be}{\begin{displaymath}}
\newcommand{\ee}{\end{displaymath}}
\newcommand{\bn}{\begin{equation}}
\newcommand{\en}{\end{equation}}
\newcommand{\mygtrsim}{\mathrel{\mbox{\raisebox{-1mm}{$\stackrel{>}{\sim}$}}}}
\newcommand{\mylsim}{\mathrel{\mbox{\raisebox{-1mm}{$\stackrel{<}{\sim}$}}}}

\newcommand{\bfm}[1]{\mbox{\boldmath$#1$}}

\newcommand{\Ni}{n_\mathrm{i}}
\newcommand{\Nn}{n_\mathrm{n}}
\newcommand{\Pii}{\bfm{\pi}_\mathrm{i}}
\newcommand{\Pin}{\bfm{\pi}_\mathrm{n}}
\newcommand{\Fi}{f_\mathrm{i}}
\newcommand{\Fgc}{f_\mathrm{i,gc}}
\newcommand{\Fn}{f_\mathrm{n}}
\newcommand{\Rn}{R_\mathrm{n}}

\usepackage[T1]{fontenc}
\usepackage{bm}
\usepackage{color}
\usepackage{graphicx}
\usepackage[breaklinks=true]{hyperref}
\hypersetup{
  bookmarks=true,         
  unicode=false,          
  pdftoolbar=true,        
  pdfmenubar=true,        
  pdffitwindow=false,     
  pdfstartview={FitH},    
  pdftitle={My title},    
  pdfauthor={Author},     
  pdfsubject={Subject},   
  pdfcreator={Creator},   
  pdfproducer={Producer}, 
  pdfkeywords={keyword1} {key2} {key3}, 
  pdfnewwindow=true,      
  colorlinks=true,        
  linkcolor=blue,         
  citecolor=blue,         
  filecolor=blue,         
  urlcolor=blue           
}

\begin{document}

\title{Plasma rotation from momentum transport by neutrals in
tokamaks}
\author{J Omotani, I Pusztai, T F\"ul\"op}
\affiliation{Department of Physics, Chalmers University of Technology,
41296 Gothenburg, Sweden} \date{\today}

\begin{abstract}
  Neutral atoms can strongly influence the intrinsic rotation and
  radial electric field at the tokamak edge. Here, we present a
  framework to investigate these effects when the neutrals dominate
  the momentum transport.  We explore the parameter space numerically,
  using highly flexible model geometries and a state of the art
  kinetic solver. We find that the most important parameters
  controlling the toroidal rotation and electric field are the major
  radius where the neutrals are localized and the plasma
  collisionality. This offers a means to influence the rotation and
  electric field by, for example, varying the radial position of the
  X-point to change the major radius of the neutral peak.
\end{abstract}

\maketitle

\renewcommand{\thesection}{\Roman{section}}
\renewcommand{\thesubsection}{\Alph{subsection}} 
The level of plasma rotation has a fundamental impact on the
performance of magnetically confined plasmas.
Establishing what determines plasma rotation is important
both from a theoretical and practical point of view, since rotation has
a strong effect on the confinement and stabilizes magnetohydrodynamic
instabilities, such as resistive wall modes.

In future magnetic fusion devices, where the effect of alpha-particle
heating will be dominant, the external torque from auxiliary heating
will be considerably lower than in current devices and the moment of
inertia will be higher. It is therefore important to understand the
intrinsic toroidal rotation that arises independently of externally
applied momentum sources; momentum transport by neutral atoms is a
mechanism that generates intrinsic rotation.

There is a wealth of experimental evidence that shows that neutrals
have a substantial influence on tokamak edge processes. They are observed to
influence global confinement \cite{Joffrin14_JET,Tamain2015} and the
transition from low (L) to high (H) confinement mode
\cite{Carreras98,Owen98,Gohil01,Boivin00,Field02_MAST,Valovic02_COMPASS,Fukuda00,Field2004,Maingi2004},
which are critical to the performance of tokamak fusion reactors.
While the physics of the transition to H-mode is far from fully
understood, it is clear that rotational flow shear plays an important
role \cite{Terry00}. It is therefore important to be able to model the effect of
neutral viscosity on the flow shear in the edge plasma.

Neutrals influence the ion dynamics in plasmas through atomic
processes, mainly through charge-exchange (CX), ionization, and
recombination.  Due to their high cross-field mobility they can be the
most significant momentum transport channel even at low relative
densities.  The effect of the neutrals is typically significant if the
neutral fraction in the plasma exceeds about $10^{-4}$ \cite{Catto98},
which is usually the case in the tokamak edge region not too far
inside the separatrix; the neutrals can penetrate even to the pedestal
top in the JET tokamak \cite{Versloot2011} and may be expected to
penetrate further in an L-mode plasma due to the lower edge density.

Recent experimental results at JET have demonstrated that changes in
divertor strike point positions are correlated with strong
modification of the global energy confinement
\cite{Joffrin14_JET,Tamain2015}. It was speculated that the reason for
this may be that neutrals are directly affecting the edge ion flow and
electric potential \cite{Joffrin14_JET}. Other experimental
observations correlate the edge intrinsic toroidal rotation with CX
dynamics \cite{Versloot2011} or the major radius of the X-point
\cite{Stoltzfus-Dueck15_PRL,*Stoltzfus-Dueck15_PoP}.  These recent
observations triggered renewed interest in the role of the neutrals in
the edge region of tokamaks.

Previous analytical work based on neoclassical theory has shown that
if the neutral contribution to the toroidal angular momentum is
dominant, neutrals can modify or even determine the edge plasma
rotation and radial electric field
\cite{Catto98,Fulop98_1,*Fulop98_2,Fulop01,Fulop02,Helander2003,Simakov2003,Dorf2013}.
It has been shown that even if there is no input of external momentum
into the plasma, the neutrals drive intrinsic momentum transport and
hence rotation.  The effect of the neutrals is sensitive to their
poloidal location which can therefore be used to control the
transport.  However, analytical solutions for the neoclassical ion
distribution function can only be obtained in the asymptotically low
or high collisionality regimes. Realistic plasma parameters in the
tokamak edge are intermediate between these two limits. The analytical
results therefore cannot describe the parametric dependence, for
experimentally relevant conditions, of the ion flow and radial
electric field or even predict the direction of the trends with, for
instance, X-point major radius. By solving the system numerically we
can remove the restriction on collisionality and are thus able to
model experimentally relevant parameters.

In this paper, we couple neutrals to a neoclassical kinetic solver,
allowing us to determine the radial electric field and plasma flows
just inside the separatrix. We use ITER-relevant model geometries and
plasma parameters. As an example application, we demonstrate the
effects of changing the X-point position and the shaping parameters,
due to their recent experimental relevance
\cite{Joffrin14_JET,Tamain2015,Stoltzfus-Dueck15_PRL,*Stoltzfus-Dueck15_PoP}.
We will show that there is a link between the X-point radial position
and neutral mediated rotation. Consequently, control of the X-point
location offers a straightforward means of external control over the
ion flow and radial electric field. Our results may have relevance for
the mechanism underlying recent observations indicating improved
global confinement with the corner divertor configurations in JET
\cite{Joffrin14_JET,Tamain2015}.

The toroidal ion flow and radial electric field can be calculated from
the steady state condition in a plasma without momentum sources, where
the radial transport of toroidal angular momentum should vanish,
\bn
\left\langle R \hat{\bfm{\zeta}} \cdot (\Pii+\Pin) \cdot \nabla \psi\right\rangle=0.
\label{1}
\en
Here $\Pii$ and $\Pin$ are the ion and neutral viscosity
tensors, $R$ is  the major radius, $2\pi\psi$ is the poloidal magnetic
flux, and $\hat{\bfm{\zeta}}=\nabla\zeta/\left|\nabla\zeta\right|$
with $\zeta$ the toroidal angle in the direction of the plasma
current.

For modest relative neutral densities, $\Nn/\Ni\mygtrsim 10^{-4}$,
where $\Nn$ and $\Ni$ denote the density of neutrals and ions, it
can be shown that the neutral viscosity is higher than the
neoclassical viscosity \cite{Catto98}. The turbulent part of the
momentum flux will also be assumed to be lower than the neutral
momentum flux, which may be questionable. However, it is plausible to
think that the neutral and turbulent momentum fluxes should be at
least comparable, given the fact that in steady state the particle
losses are balanced by fueling and recycling. The neutral particle
transport is then equal to ion particle transport due to both
collisions {\em and} turbulence, since every recycling ion that leaves
the plasma comes back as a neutral \cite{Helander2003}. Therefore
considering the neutral momentum transport in isolation is an
important step in understanding plasma rotation in regions where
neutrals are present.  The importance of neutrals for rotation is
reinforced by the experimental evidence showing that neutrals do
affect plasma rotation at the edge.

We can solve the neutral kinetic equation $\g v \cdot \nabla \Fn =
\tau^{-1} \lp \Nn\Fi/\Ni-\Fn\rp$ perturbatively. Here, $\tau^{-1}= \Ni
\langle \sigma v\rangle_x \simeq 2.93 \Ni \sigma_x(T_i/m_i)^{1/2}$ is
the CX frequency, which is much larger than the ionization or
recombination rates for tokamak edge parameters. The mean free path
for CX is $\lambda_\mathrm{mfp,n}=\tau v_\mathrm{th}\simeq
0.483/\Ni\sigma_x$, with $v_\mathrm{th}$ the thermal velocity, which
we may estimate as $\lambda_\mathrm{mfp}\simeq 0.8\;\mathrm{cm}$ using
$\Ni=10^{20}\;\mathrm{m}^{-3}$ and $\sigma_x=6\times
10^{-15}\;\mathrm{cm}^2$ \cite{Catto98}. This is short compared to
typical gradient scale lengths in the plasma, so we expand the neutral
distribution function for small $\lambda_\mathrm{mfp}/L$, where $L$ is
a typical gradient scale length, as
$\Fn=f_\mathrm{n0}+f_\mathrm{n1}+\ldots$.  To lowest order
$f_\mathrm{n0}=\Nn \Fi/\Ni$, and to next order $f_\mathrm{n1}=-\tau \g
v \cdot \nabla\left(\Nn \Fi/\Ni\right)$. Thus the neutral distribution
function can be calculated from the distribution function of the ions.
For neutral fractions $\Nn/\Ni\mylsim 10^{-3}$, the direct effect of
the neutrals on the ion distribution function can be neglected
\cite{Fulop98_1,*Fulop98_2}, and then we can construct the neutral
viscosity tensor as $\pi_{\mathrm{n},jk}=\int m_\mathrm{i}\lp v_j
v_k-(v^2/3)\delta_{jk}\rp \Fn(\g v) d^3 v=-\tau \frac{\p}{\p x_l}\int
m_\mathrm{i} v_j v_k v_l (\Nn/\Ni)\Fi (\g v)d^3
v+(\ldots)\pi_{\mathrm{i},jk}+(\ldots)\delta_{jk}$, where the last two
terms are negligible compared to the first one, so we have
\begin{align} &\left<R\hat{\boldsymbol\zeta}\cdot\boldsymbol
\pi_\mathrm{n} \cdot\nabla\psi\right> \nonumber\\ &\approx
\left<\frac{R \tau
m_\mathrm{i}}{\Ni}\,\frac{\partial\Nn}{\partial\psi} \int
d^3v\,\left(\nabla\psi\cdot\boldsymbol v
\vphantom{\hat{\boldsymbol\zeta}}\right)^2\left(\hat{\boldsymbol\zeta}\cdot\boldsymbol
v\right)\Fi\right>,\label{2} \end{align} where the radial gradient of
$\Nn$ dominates.  Note that even when the neutrals are too few to
affect the ion distribution, they still affect the ion toroidal
rotation and the radial electric field through the transport of
angular momentum. Interestingly, from Eqs.~(\ref{1}) and (\ref{2}) it
follows that the effect does not depend on the magnitude of the
neutral density or on the cross section if the neutral viscosity is
larger than the ion neoclassical and turbulent viscosities, as long as
there are few enough neutrals not to affect the ion distribution
function directly.

We calculate the guiding-center ion distribution function $\Fgc$ with
the Pedestal and Edge Radially-global Fokker-Planck Evaluation of
Collisional Transport (\textsc{perfect}) neoclassical solver
\cite{Landreman14_PERFECT}, used here in local mode. In contrast to
standard neoclassical calculations which need consider only the
guiding center distribution $\Fgc$, the neutrals couple to the
\emph{particle} distribution $\Fi$, given by \begin{align}
  \Fi(\boldsymbol r) &= \Fgc^\mathrm{M}(\boldsymbol R) +
  \delta\Fgc(\boldsymbol R) \nonumber\\ &\approx
  \Fgc^\mathrm{M}(\boldsymbol r) -
  \boldsymbol\rho\cdot\nabla\Fgc^\mathrm{M}(\boldsymbol r)
+\delta\Fgc(\boldsymbol r).\label{3} \end{align} Here $\boldsymbol r$
is the particle position, $\boldsymbol R$ the guiding center position,
$\boldsymbol \rho = \boldsymbol R-\boldsymbol r$ the gyro-radius
vector, and subscript `gc' denotes guiding center distributions.  We
model here deuterium ions and neutrals (electron dynamics are
negligible due to the small electron-ion mass ratio) and iterate to
find the radial electric field which makes the momentum flux due to
neutrals vanish.

In the case of the local drift kinetic equation (as defined, for
example, in Ref.~\cite{Landreman14_PERFECT}), the system is linear;
further, once we solve for the self-consistent electric field the
toroidal velocity is proportional to the temperature gradient.  The
reason for this is that to solve the drift kinetic equation for
$\delta f$, the inputs on the right hand side of the equation are the
gradients of density, temperature, and electrostatic potential $\Phi$.
Since the $\partial\Ni/\partial\psi$ and $\partial\Phi/\partial\psi$
terms have identical velocity space structures, any density gradient
is just offset by $\partial\Phi/\partial\psi$ without affecting
$\delta f$ and hence without affecting the flow.  Since the only
effect of the density gradient is to give a constant offset to the
electric field, we set $\partial\Ni/\partial\psi=0$. Since we solve
for the electric field, the only driving term is the ion temperature
gradient.  The system is thus fully specified and so the toroidal
rotation is determined.

\begin{figure}[tbp]
  \includegraphics[height=78pt]{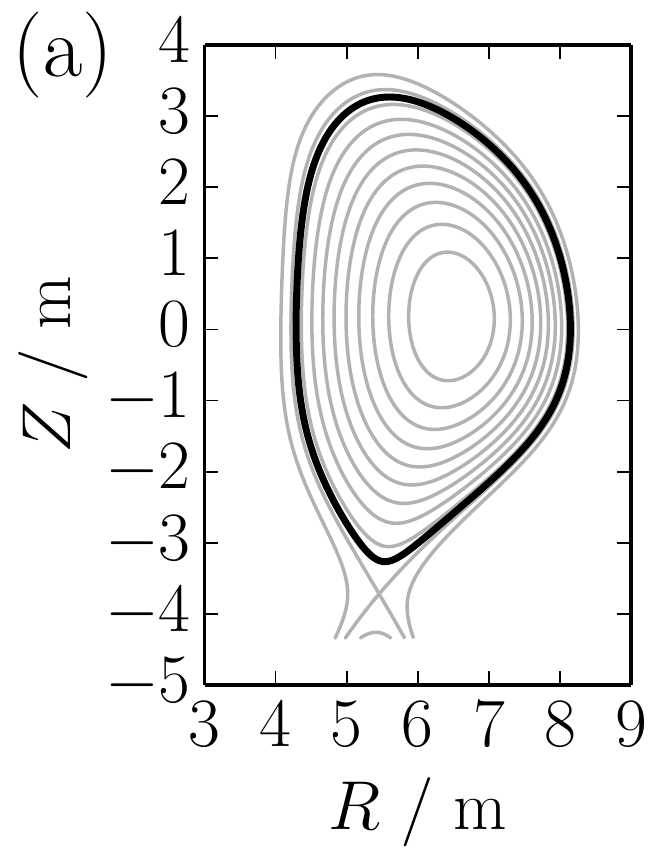}
  \includegraphics[height=78pt]{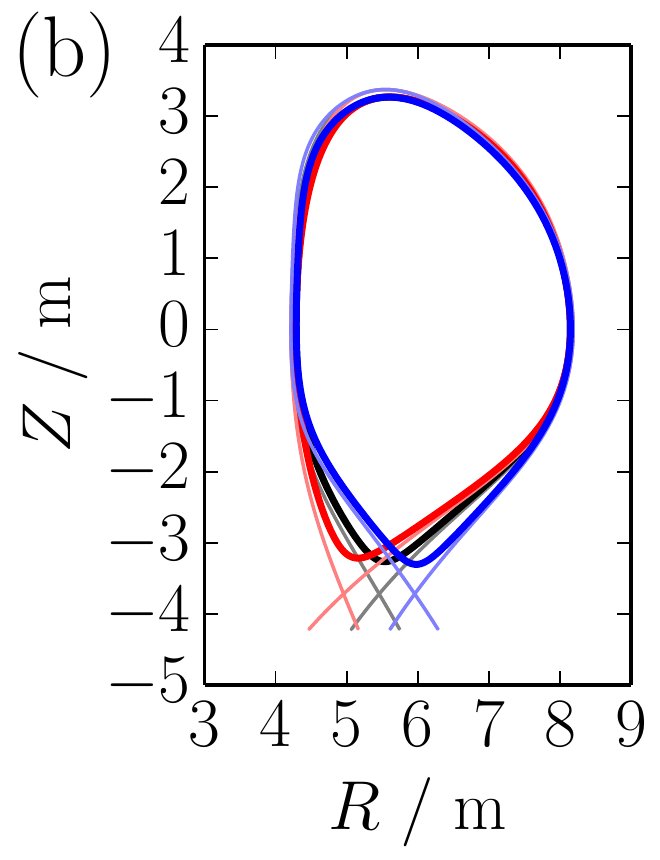}
  \includegraphics[height=78pt]{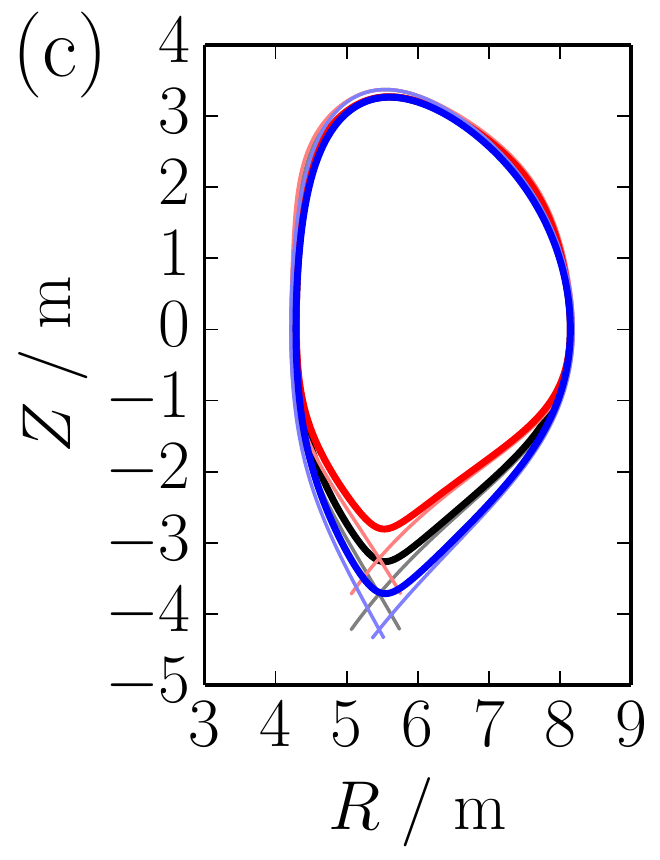}
  \includegraphics[height=78pt]{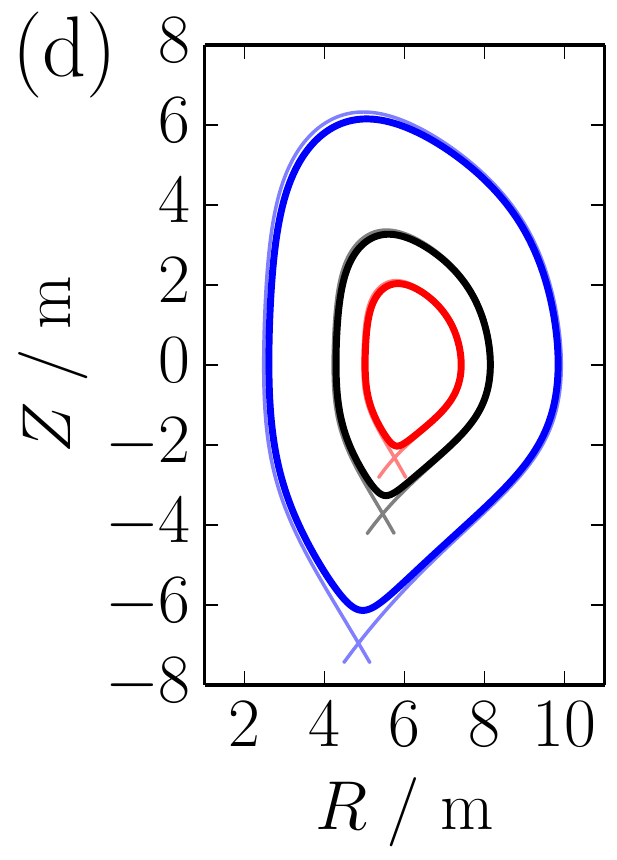} \caption{Flux
    surface shapes: (a) baseline ITER equilibrium, (b) changing the
    major radius of the X-point $R_\mathrm{X}$, (c) changing the
    vertical position of the X-point $Z_\mathrm{X}$, and (d) changing
the inverse aspect ratio $\epsilon$. Thick lines show the
$\psi_\mathrm{N}=0.95$ surface used for simulations.}\label{fl}
\end{figure}

We explore the effects of X-point position and shaping parameters,
using model magnetic geometries given by analytic solutions to the
Grad-Shafranov equation \cite{Cerfon10}, with ITER-like parameters.
These analytical solutions allow for arbitrary inverse aspect ratio
$\epsilon$, elongation $\kappa$, triangularity $\delta$, and X-point
position. The geometry is specified by fixing the shape of the
boundary surface in terms of $\epsilon$, $\kappa$, $\delta$, and
X-point position major radius $R_\mathrm{X}$ and height
$Z_\mathrm{X}$. Two more constraints are also needed. We take the
toroidal $\beta$, $2 \mu_0 \langle p\rangle/B_0^2$ (where $\langle
p\rangle$ is the volume averaged pressure), to be 0.05 and fix the
safety factor $q_{95}$ of the flux surface with
$\psi_\mathrm{N}=0.95$, where $\psi_\mathrm{N}$ is the normalized
poloidal flux, to the value that corresponds to a plasma current of 15
MA in the baseline equilibrium.  Scales are set by giving $R_0$, the
major radius of the plasma center, and $B_0$, the vacuum toroidal
field at $R_0$.  The baseline parameters are \cite{Cerfon10}:
$R_0=6.2\mathrm{m}$, $B_0=5.3\mathrm{T}$, $\epsilon=0.32$,
$\kappa=1.7$, $\delta=0.33$, $R_\mathrm{X}=(1-1.1\delta\epsilon)R_0$,
$Z_\mathrm{X}=-1.1\kappa\epsilon R_0$. The variation of the X-point
position and inverse aspect ratio, while keeping the other shaping
parameters fixed, is illustrated in Fig.~\ref{fl}.

Clearly there is a rich parameter space of edge physics to explore
with this numerical approach, with the plasma collisionality, the
spatial distribution of the neutral density and the magnetic geometry
all influencing the solutions. Our numerical tool allows  these
effects to be taken into account simultaneously. We also find good
agreement with analytical results \cite{Fulop02,Helander2003} in the
appropriate limits.

It is interesting to investigate the consequences of varying the
position of a localized concentration of neutrals. The neutrals may be
localized at a particular poloidal position, representing the location
of a gas puff. They may also be concentrated near the X-point when the
recycling from the targets is strong \cite{Callen2010,Versloot2011},
or if the plasma is gas fueled from the private flux region. Therefore
we consider two scenarios. Firstly we vary the poloidal location of
the neutrals in the baseline geometry and secondly  vary the geometry
in various ways while keeping the neutrals fixed at the X-point. In
all cases we consider a single flux surface at $\psi_\mathrm{N}=0.95$,
as highlighted with thick lines in Fig.~\ref{fl}.

\begin{figure}[tb]
  \includegraphics[width=0.352\textwidth]{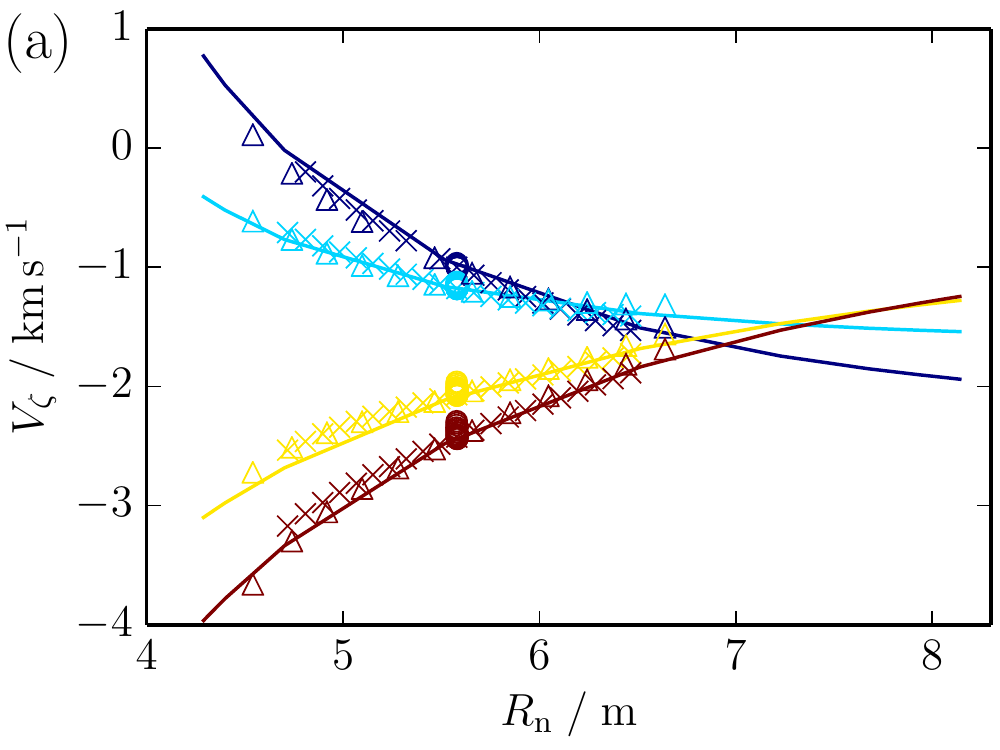}

  \includegraphics[width=0.352\textwidth]{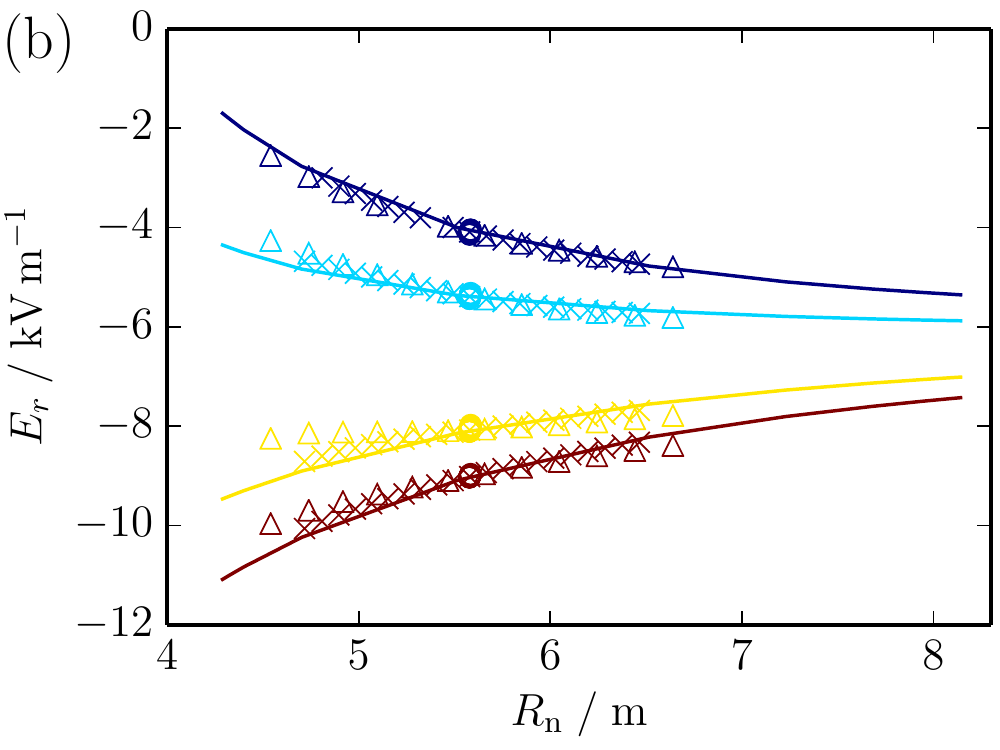} \caption{(a)
    Toroidal flow velocity and (b) radial electric field at the
    outboard midplane as a function of the major radius where the
    neutrals are localized. Lines show the effect of changing the
    poloidal position of neutrals in the baseline geometry.  Markers
    show the effect, with neutrals kept fixed at the X-point, of
    changing the geometrical parameters $R_\mathrm{X}$ (crosses),
    $Z_\mathrm{X}$ (circles), and $\delta$ (triangles).  Line colors
    correspond to collisionality: cyan is the baseline corresponding
    to $\Ni=10^{20}\;\mathrm{m}^{-3}$ and
$T_\mathrm{i}=300\;\mathrm{eV}$, blue is 10 times lower, and yellow
and red are 10 and 100 times higher respectively. The ion temperature
scale length is taken to be 10 cm.} \label{fig:V-overlay} \end{figure}

The results show that the toroidal flow and electric field are largely
determined by the major radius where the neutrals are localized,
$\Rn$, and the plasma collisionality. This is illustrated in
Fig.~\ref{fig:V-overlay}, where we show the toroidal ion flow and
radial electric field as a function of the major radius where the
neutrals were located, for various collisionalities.  We show scans of
the major radius of the X-point $R_X$, height of the X-point $Z_X$,
and triangularity $\delta$, all of which have the neutrals located at
the X-point. We also scanned the poloidal location $\theta$ of the
neutrals in the baseline equilibrium ($\theta=0$ corresponding to the
largest $\Rn$, at the outboard midplane, and $\theta=\pi$ to the
smallest, at the inboard midplane). Since the scans in $\theta$,
$R_\mathrm{X}$, $Z_\mathrm{X}$, and $\delta$ collapse on a single
curve, we can see that the position of the X-point and triangularity
affect the flow only by changing $\Rn$, while the details of changes
to the flux surface geometry are much less significant. One way to
imagine controlling the major radius where the neutrals are localized
is by changing the major radius of the X-point, and another is with
the fueling location.

\begin{figure}[tb]
  \includegraphics[width=0.22\textwidth]{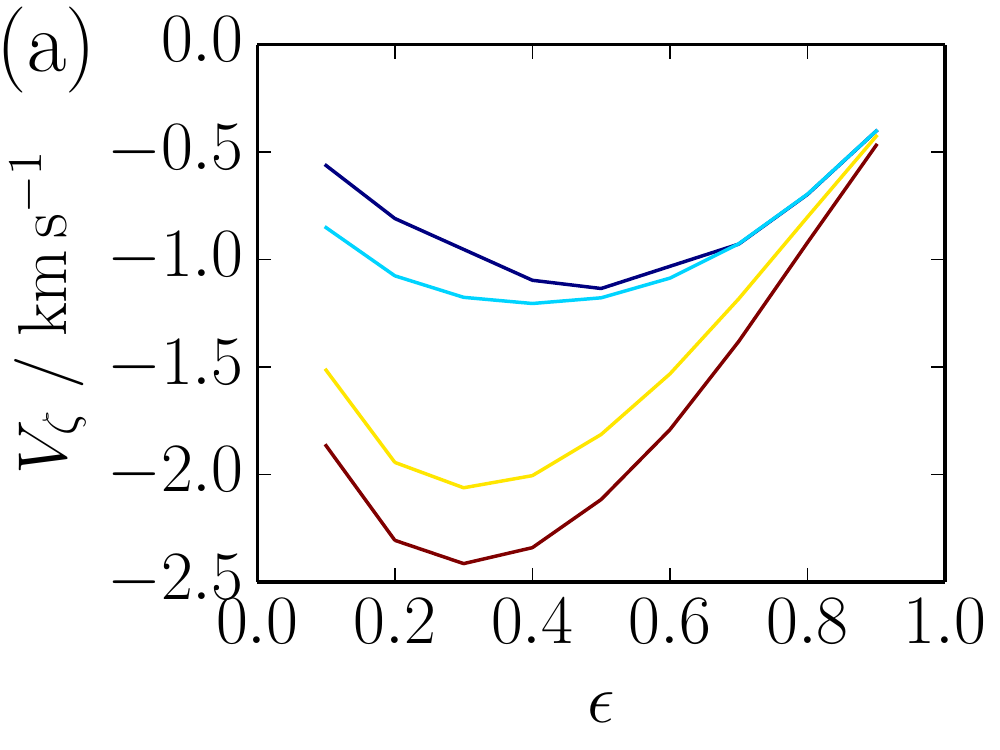}
  \includegraphics[width=0.22\textwidth]{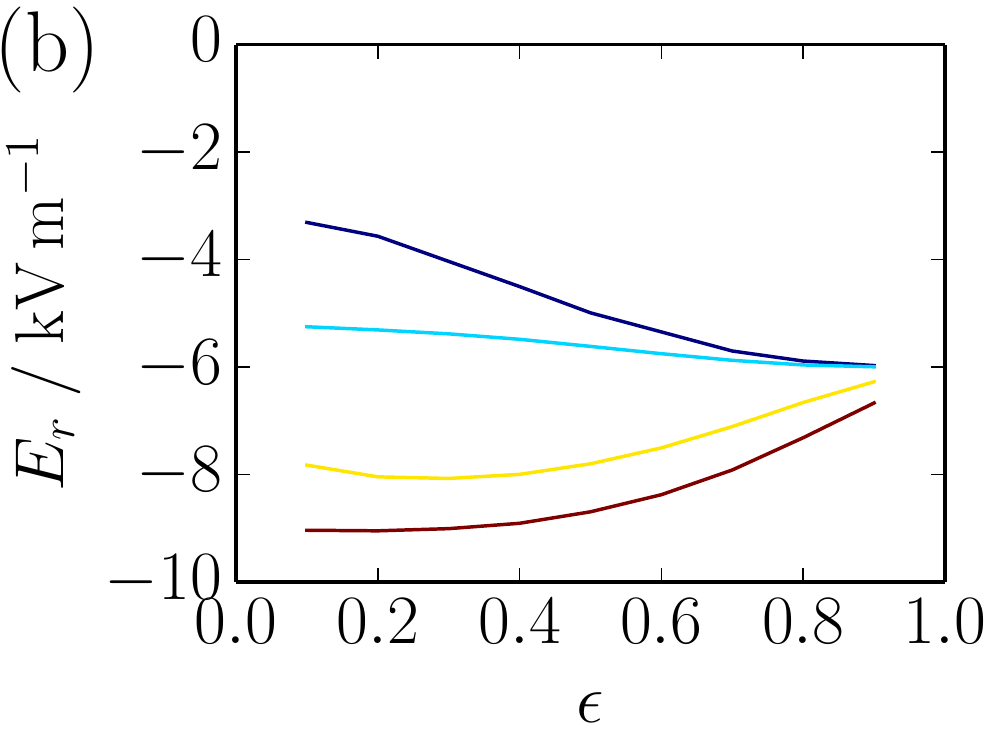}

  \includegraphics[width=0.22\textwidth]{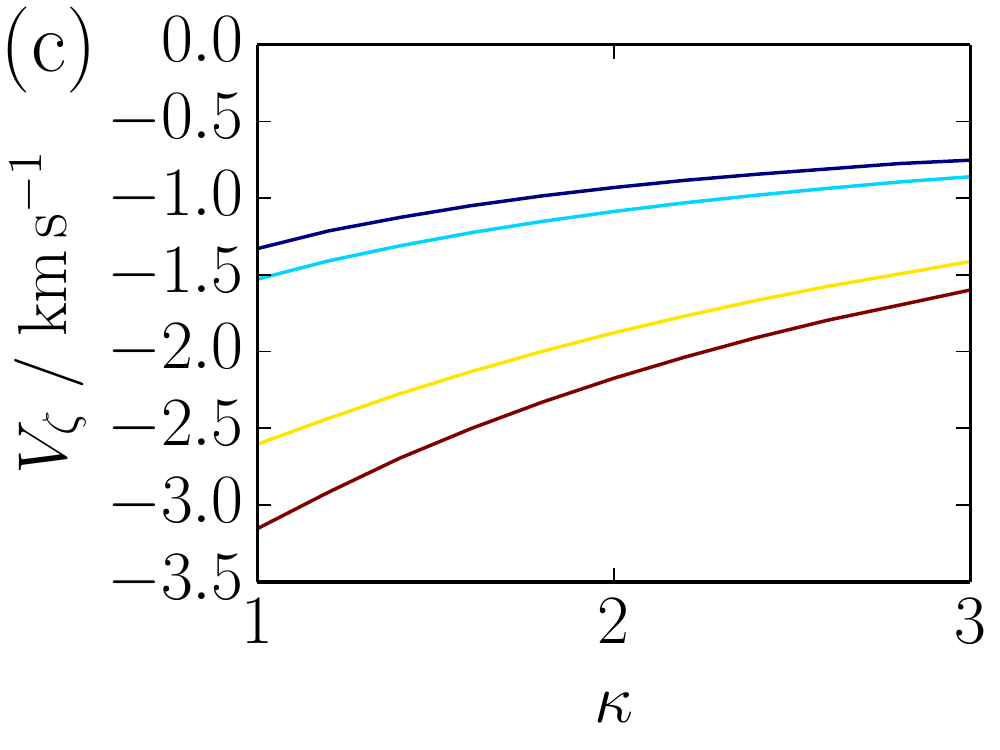}
  \includegraphics[width=0.22\textwidth]{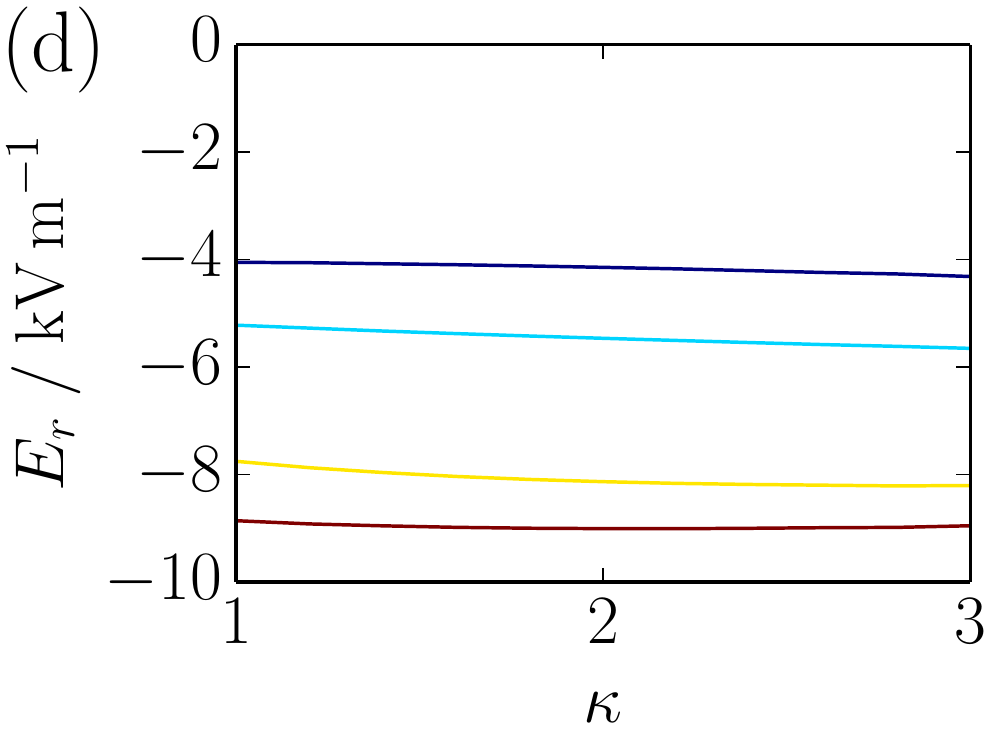}
  \caption{(a) Toroidal flow velocity and (b) radial
    electric field at the outboard midplane for several inverse aspect
    ratios $\epsilon$.  (c) and (d) the same for elongations $\kappa$.
    Line colors correspond to collisionalities as in
Fig.~\ref{fig:V-overlay}. The ion temperature scale length is again
taken to be 10 cm.} \label{fig:epsilon-kappa} \end{figure}

In contrast, the lowest order shaping parameters (inverse aspect ratio
$\epsilon$ and elongation $\kappa$) have an effect on the flow and
electric field that is not just described by $\Rn$ as in
Fig.~\ref{fig:V-overlay}.  Figure \ref{fig:epsilon-kappa} shows that
as $\epsilon$ varies there is an extremum in the toroidal flow. The
collisionality dependence of both flow and electric field is
suppressed for large $\epsilon$. $\kappa$ does not affect the electric
field much, but the toroidal rotation decreases in magnitude as
$\kappa$ increases, as shown in Fig.~\ref{fig:epsilon-kappa}.

The neutrals cause the plasma to rotate toroidally even in the absence
of external momentum input. This is due to the fact that toroidal flow
is needed to give a radial momentum flux balancing that due to the
toroidal heat flux, which is present because of the radial temperature
gradient \cite{Helander2003}. In the limits $\Rn\rightarrow\infty$ in
Fig.~\ref{fig:V-overlay} and $\epsilon\rightarrow 1$ in
Fig.~\ref{fig:epsilon-kappa} the electric field $E_r$ approaches
$-6\;\mathrm{kV}\;\mathrm{m}^{-1}=-2T_i/(e L_{T_i})$, where $L_{T_i}$
is the gradient scale length of the ion temperature. This is because
in these limits only the rigid rotation parts of the flow and heat
flux 
contribute to drive the  radial momentum flux and these depend only on
the radial gradients, not on $\delta \Fgc$; they are therefore
independent of collisionality and so is the electric field which is
set directly. The plasma flow in the $\Rn\rightarrow\infty$ limit does
also have a contribution from the part of the flow parallel to the
magnetic field (the part governed by the neoclassical coefficient $k$)
and so does depend slightly on collisionality. It does not contribute
to the momentum flux because $B(\Rn)\rightarrow0$ in this limit. It is
also notable that the trends in $V_\zeta$ and $E_r$ with $\Rn$ (Fig.
\ref{fig:V-overlay}) reverse their direction as the collisionality
changes from low to high. This follows from the change in sign of the
neoclassical flow coefficient $k$ between low (banana) and high
(Pfirsch-Schl\"uter) collisionalities.

We find that the toroidal flow caused by the neutrals is generally
counter-current and stronger for higher collisionality. The magnitude
of the rotation in JET L-mode plasmas without external torque (without
NBI heating) may be a few $\mathrm{krad}\;\mathrm{s}^{-1}$,
corresponding to outboard velocities of order
$10\;\mathrm{km}\;\mathrm{s}^{-1}$ at a major radius of
$3.7\;\mathrm{m}$ \cite{Eriksson2009}; thus the speeds of a few
$\mathrm{km}\;\mathrm{s}^{-1}$ found here are of the same order of
magnitude and at least likely to compete with other effects driving
intrinsic edge rotation. The electric field is always inwards and is
also stronger for higher collisionality. The effect of collisionality
is enhanced when the neutrals are located at smaller major radii.

Allowing higher neutral densities requires including the reaction of
the ion distribution to the neutrals \cite{Fulop98_1}, which will be
implemented numerically in the future.  \textsc{perfect} has the
capability to include finite orbit width effects
\cite{Landreman14_PERFECT,Pusztai2016}, allowing density pedestals to
be modeled and the study of the interaction between neutral momentum
transport and pedestals is of the highest importance.  The importance
of and interaction with other effects such as ion orbit loss
\cite{Dorf2016,Stoltzfus-Dueck12_PRL,*Stoltzfus-Dueck12_PoP} could
also be considered.  

{\em Summary.}  We have built a framework to investigate the toroidal
rotation and radial electric field in the edge plasma, when these are
regulated through momentum transport by neutrals, by coupling neutrals
to a neoclassical kinetic solver. Experimentally relevant parameters
are not described by the asymptotic collisionality limits that can be
studied analytically
\cite{Fulop98_1,*Fulop98_2,Fulop01,Fulop02,Helander2003,Simakov2003},
which cannot, for intermediate collisionality, predict even the
qualitative trends that should be expected. Intermediate
collisionality is typical of experiments, see the cyan curves for
baseline parameters in Figs.\ \ref{fig:V-overlay} and
\ref{fig:epsilon-kappa}.  Therefore quantitative comparison with
experiment and predictive power for future devices both require the
numerical solutions that we present here.

We find that the most important parameters that control the toroidal
flow and electric field are the major radius where the neutrals are
localized, $\Rn$, and the plasma collisionality. These results suggest
that altering the X-point position may offer a means to manipulate the
edge rotation in the layer inside the separatrix where neutral
viscosity dominates.  This sets the boundary condition for the core
rotation profile and influences the stability of magnetohydrodynamic
instabilities such as resistive wall modes.  Further, shear in the
edge rotation can lead to the suppression of edge turbulence.
Consequently the neutrals are also likely to affect the L-H transition
and H-mode confinement. Our results demonstrate that the effects of
neutrals on momentum transport are significant and should be accounted
for both in the interpretation of current experiments and in the
design of future machines.  

The authors are grateful to Sarah Newton for useful discussions and to
Matt Landreman for advice and help with the \textsc{perfect} code.
This work was supported by the Framework grant for Strategic Energy
Research (Dnr.~2014-5392) and the International Career Grant
(Dnr.~330-2014-6313) from Vetenskapsr{\aa}det.

\bibliography{references}

\end{document}